\documentclass[aps,prb,superscriptaddress,longbibliography,floatfix,preprint,showpacs]{revtex4-1}

\usepackage[pdftex]{graphicx}
\usepackage{bm}
\usepackage[czech,english]{babel}	

\usepackage[pdftex,colorlinks,linkcolor=blue,urlcolor=blue,citecolor=blue,bookmarks=false]{hyperref}
\usepackage{color}

\begin{document}

\title{Imaging ambipolar two-dimensional carriers induced by the spontaneous electric polarization of a polar semiconductor BiTeI}

\newcommand{\CEMS}{RIKEN Center for Emergent Matter Science, Wako, Saitama 351-0198, Japan.}
\newcommand{\ASI}{RIKEN Advanced Science Institute, Wako, Saitama 351-0198, Japan.}
\newcommand{\MSL}{Materials and Structures Laboratory, Tokyo Institute of Technology, Yokohama, Kanagawa 226-8503, Japan.}

\author{Y. Kohsaka}
\email{kohsaka@riken.jp}
\affiliation{\CEMS}
\affiliation{\ASI}

\author{M. Kanou}
\affiliation{\MSL}

\author{H. Takagi}
\affiliation{\ASI}
\affiliation{RIKEN Magnetic Materials Laboratory, Wako, Saitama 351-0198, Japan.}
\affiliation{Department of Physics, University of Tokyo, Bunkyo-ku, Tokyo 113-0033, Japan.}
\affiliation{Max-Planck-Institut f\"ur Festk\"orperforschung, Heisenbergstra\ss e 1 70569 Stuttgart, Germany.}

\author{T. Hanaguri}
\affiliation{\CEMS}

\author{T. Sasagawa}
\affiliation{\MSL}

\date{\today}

\pacs{73.20.At, 68.37.Ef, 71.70.Ej}

\begin{abstract}
Two-dimensional (2D) mobile carriers are a wellspring of quantum phenomena.
Among various 2D-carrier systems, such as field effect transistors and heterostructures, polar materials hold a unique potential;
the spontaneous electric polarization in the bulk could generate positive and negative 2D carriers at the surface.
Although several experiments have shown ambipolar carriers at the surface of a polar semiconductor BiTeI, their origin is yet to be specified.
Here we provide compelling experimental evidences that the ambipolar 2D carriers at the surface of BiTeI are induced by the spontaneous electric polarization.
By imaging electron standing waves with spectroscopic imaging scanning tunneling microscopy, we find that positive or negative carriers with Rashba-type spin splitting emerge at the surface correspondingly to the polar directions in the bulk.
The electron densities at the surface are constant independently of those in the bulk, corroborating that the 2D carriers are induced by the spontaneous electric polarization.
We also successfully image that lateral $p$-$n$ junctions are formed along the boundaries of submicron-scale domains with opposite polar directions. Our study presents a novel means to endow non-volatile, spin-polarized, and ambipolar 2D carriers as well as, without elaborate fabrication, lateral $p$-$n$ junctions of those carriers at atomically-sharp interfaces.
\end{abstract}

\maketitle

\section{Introduction}

The spontaneous electric polarization of polar materials causes surface charges, or band bending near the surface.
If the band bending is large enough for the conduction and valence bands to cross the Fermi level, two-dimensional (2D) positive carriers can emerge at one side of a material and negative ones at the other side.
Such mobile carriers, if realized, expand an arena of 2D-carrier systems that exhibit a wide variety of quantum phenomena~\cite{Ando1982, Hwang2012}.
A pioneering work on BaTiO$_3$ claimed that the electric conductivity was increased by the polarization~\cite{Watanabe2008}.
However, carrier polarities were not shown and thus their relationship with the polarization is unclear.
To clarify the relationship, it is necessary to carefully examine surface modifications, as emphasized in Ref.~\onlinecite{Watanabe2008}.
For example, oxygen vacancies created on the fractured surface produce a 2D electron gas at the surface of SrTiO$_3$ (Ref.~\onlinecite{Santander-Syro2011, Wang2014}).
In contrast to the simple mechanism, it has never been established whether the spontaneous electric polarization actually induces 2D carriers.

A polar semiconductor BiTeI is an emergent candidate possessing the polarization-induced 2D carriers. 
BiTeI has a layered crystal structure with triple layers composed of Te, Bi, and I layers stacking along the $c$-axis~\cite{Tomokiyo1977, Shevelkov1995, Kanou2013}, as shown schematically in Fig.~\ref{fig:BiTeI}(a).
In this polar crystal structure, the spin degeneracy in the band structures is lifted by spin-orbit interaction~\cite{Rashba1960}.
Angle-resolved photoemission spectroscopy (ARPES) reveals that BiTeI indeed exhibits momentum-dependent (Rashba-type) spin splitting larger than ever reported~\cite{Ishizaka2011, Crepaldi2012, Landolt2012, Sakano2013, Sakano2012, Crepaldi2014}.
Because of this feature, BiTeI is proposed as a component for spintronics applications~\cite{Tsutsui2012}.
In addition, both $n$-type and $p$-type two-dimensional (2D) band dispersions are observed at the surface of BiTeI~\cite{Crepaldi2012, Landolt2012}.
Although preceding studies have shown the presence of band bending near the surface accompanied with the surface carriers~\cite{Ishizaka2011, Crepaldi2012, Landolt2012, Eremeev2012}, the origin of the band bending (and thus the origin of the surface carriers) remains elusive.
Ishizaka {\it el al.} indicate similarity to near-surface electron-accumulation layers of semiconductors~\cite{Ishizaka2011}.
Eremeev {\it et al.} suggest that the breaking translational symmetry at the surface with the strong ionicity modifies the electrostatic potential near the surface~\cite{Eremeev2012}.
In contrast to these surface origins, Butler {\it et al.} speculate that the spontaneous electric polarization in the bulk causes the spectral shift to discuss the surface structures~\cite{Butler2014}.
Actually, the surface structure necessary to identify the origin is also still obscure.
Multiple termination layers hosting the ambipolar carriers are attributed to stacking faults (an excess or deficiency of an atomic layer) and steps~\cite{Crepaldi2012}, or opposite stacking sequences~\cite{Landolt2012, Tournier-Colletta2014, Butler2014}.
Surface vacancies or absorbates have not been addressed in the preceding microscopy~\cite{Tournier-Colletta2014, Butler2014}.

In this study, we substantiate that the ambipolar 2D carriers at the surface of BiTeI are induced by the spontaneous electric polarization.
A core challenge, defined by the previous studies, is to elucidate local electronic states including carrier polarities, together with surface structures from submicron- to atomic- scales.
For this purpose, we performed spectroscopic-imaging scanning tunneling microscopy (SI-STM) which yields images of the local density of states (LDOS) by measuring spatial variation of the differential tunneling conductance, d$I$/d$V$.
To determine local carrier polarity, we exploit electron standing waves appearing as spatial modulations in d$I$/d$V$ images~\cite{Hasegawa1993, Crommie1993}.
An electron standing wave is caused by quantum interference between electron waves incident to and elastically scattered from an atomic defect or step.
The wavevector of electron standing wave is given as difference between those of original electron waves, $\bm{q} = \bm{k}_s-\bm{k}_i$.
Therefore, the dispersion relationship $\bm{q}(V)$, specifically the sign of d$|\bm{q}|$/d$V$, directly reflects carrier polarity: negative for holes and positive for electrons.

\section{Methods}

Single crystals of pristine and substituted BiTeI used in this study were grown by a modified Bridgman method~\cite{Kanou2013}.
All samples are doped with electrons due to non-stoichiometry.
Bulk electron densities are determined by the Hall coefficient measured at room temperature.
The $10^{19}$ cm$^{-3}$ samples are metal with temperature-independent electron densities~\cite{Kanou2013,Horak1981}.
The Fermi levels of $10^{19}$ cm$^{-3}$ samples lie $\sim$0.1 eV above the bulk conduction band minimum~\cite{Lee2011,Murakawa2013,Wang2013}.
Meanwhile, since the $10^{17}$ cm$^{-3}$ samples are not fully metallic and have the Fermi levels near the bulk conduction band minimum, their electron densities at low temperatures where SI-STM measurements were performed may be smaller than those at room temperature.
Therefore, the 0.1 eV difference of the Fermi levels and the two orders of magnitude difference of the bulk electron densities are the minimum estimates of those at low temperatures.

For SI-STM measurements, BiTeI crystals were cleaved in an ultra-high vacuum chamber at $\sim$77~K.
The crystals were then immediately transferred with a transfer rod cooled together with the crystals, through an insert cooled by liquid helium, to a home-built STM head placed at the bottom of the insert and cooled down to 4.6~K beforehand.
All SI-STM measurements were carried out at 4.6~K with tungsten tips sharpened electrochemically and prepared with a field ion microscope.
Bias voltages were applied to the sample and the tip was virtually grounded.
Topographic images were taken in the constant-current mode. d$I$/d$V$ spectra and images were measured by a standard lock-in technique with a setup current of 0.2~nA and the feedback loop opened.
The modulation voltage was 5~mV$_\mathrm{rms}$ unless otherwise noted.
Fourier transforms of d$I$/d$V$ images are symmetrized based on $C_\mathrm{3v}$ symmetry of BiTeI.

\section{Results}

\begin{figure*}
	\includegraphics{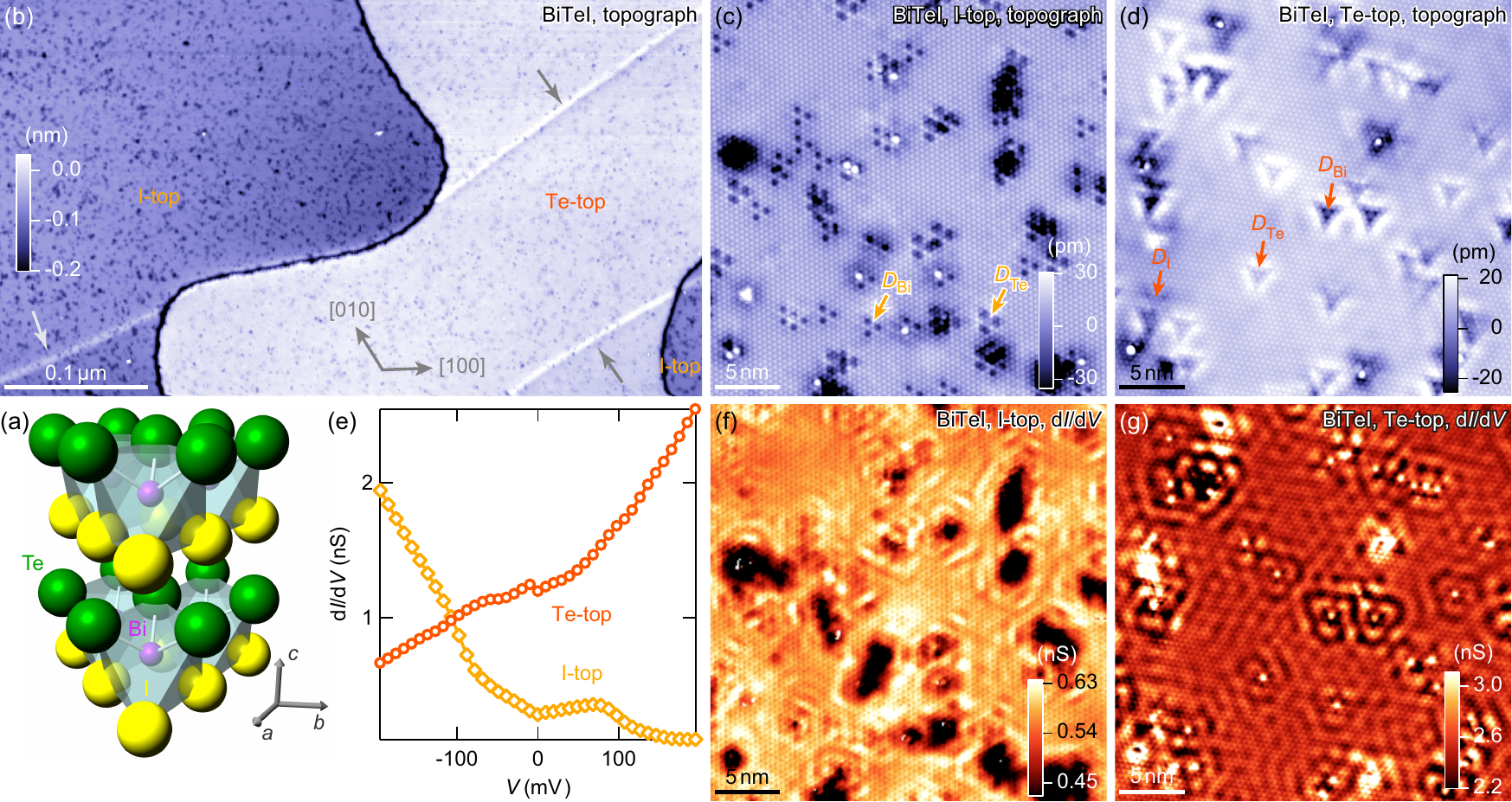}
	\caption{
		(Color online)
		Two types of domains of BiTeI.
		(a) Crystal structure of BiTeI. 
		(b) A 0.49~$\mu$m $\times$ 0.28~$\mu$m topographic image.
		The scanning parameters are 10~pA at -0.2~V.
		The gray arrows with indices show the lattice orientation determined by the Bragg peaks in Fourier transforms.
		Filamentary structures indicated by the other arrows are only occasionally observed and irrelevant to our arguments.
		(c and d) 30 $\times$ 30~nm$^2$ topographic images of each domain.
		The scanning parameters are 0.2~nA at -0.2~V for (c), and 0.2~nA at -20~mV for (d).
		The arrows indicate examples of defects.
		The annotation $D_\mathrm{X}$ denote defects at X-site (X = Bi, Te, and I).
		Dark patches in (c) are clusters of some defects.
		(e) d$I$/d$V$ spectra averaged in each domain. 
		The setup bias voltage is -0.2~V.
		(f and g) d$I$/d$V$ images taken in the same areas as (c) and (d), respectively.
		(f) was taken at -195~mV and (g) at +195~mV.
		The setup bias voltages are -0.39~V for (f) and -0.2~V for (g).
	\label{fig:BiTeI}}
\end{figure*}

To grasp submicron-scale characteristics of BiTeI, we show a large topographic image in Fig.~\ref{fig:BiTeI}(b).
A prominent structure observed in all samples studied is the two types of domains identified topographically and electronically as described below.
The domains are typically several hundred nanometers in size and separated by depression in topographic images.
The domain boundaries are not straight along the crystalline axes but meandering smoothly.
One type of the domains is apparently higher than the other by about 0.1~nm, which weakly depends on bias voltages (Appendix~\ref{appendix:relativeheight}).
The bias dependence means that the topmost layers of the domains are crystallographically inequivalent.

A topographic distinction between the two types of domains is defect patterns appearing in high-resolution topographic images shown in Fig.~\ref{fig:BiTeI}(c) and \ref{fig:BiTeI}(d).
Besides the triangular lattices with the bulk $a$-axis constant, atomic defects are imaged as three dots in Fig.~\ref{fig:BiTeI}(c), but dark triangles in Fig.~\ref{fig:BiTeI}(d).
The fact that only two kinds of domains are observed indicates that the domains are composed of opposite stacking sequences, I-Bi-Te (I-top) and Te-Bi-I (Te-top).
(See Appendix~\ref{appendix:stacking} for detailed description.)
More details about the domain structure are brought by further investigating topographic images.
First, stacking sequences of each domain are I-top for Fig.~\ref{fig:BiTeI}(c) and Te-top for Fig.~\ref{fig:BiTeI}(d).
Second, the structural relationship between the domains is $\overline{2}$ (reflection about the (0001) plane).
Third, central sites of defect patterns are identified as shown in Fig.~\ref{fig:BiTeI}(c) and \ref{fig:BiTeI}(d).
(See Appendix~\ref{appendix:domaindetails} for full descriptions.)
These thorough identifications of the domain structure play an important role to understand electronic states of each domain as described below.

An electronic distinction between the two types of domains is represented by d$I$/d$V$ spectra shown in Fig.~\ref{fig:BiTeI}(e).
The spectrum of the Te-top domain shows finite conductance in the voltage range studied whereas that of the I-top domain shows vanishingly small conductance at positive bias voltages.
The latter implies that, at the surface of the I-top domain, the top of valence band is located slightly above the Fermi level and thus the charge carriers are holes in contrast to electrons in the bulk.
d$I$/d$V$ images also clearly manifest the distinction as shown in Fig.~\ref{fig:BiTeI}(f) and \ref{fig:BiTeI}(g).
Electron standing waves are observed mainly near the defects on the I-top domain while all over on the Te-top domain~\cite{SMdIdVmovie}.
By comparing these d$I$/d$V$ images with the topographic images taken in the same location (Fig.~\ref{fig:BiTeI}(c) and \ref{fig:BiTeI}(d)), we find that strong scattering centers are defects at Te-site in the I-top domain and Bi-sites in the Te-top domain.

\begin{figure*}
	\includegraphics{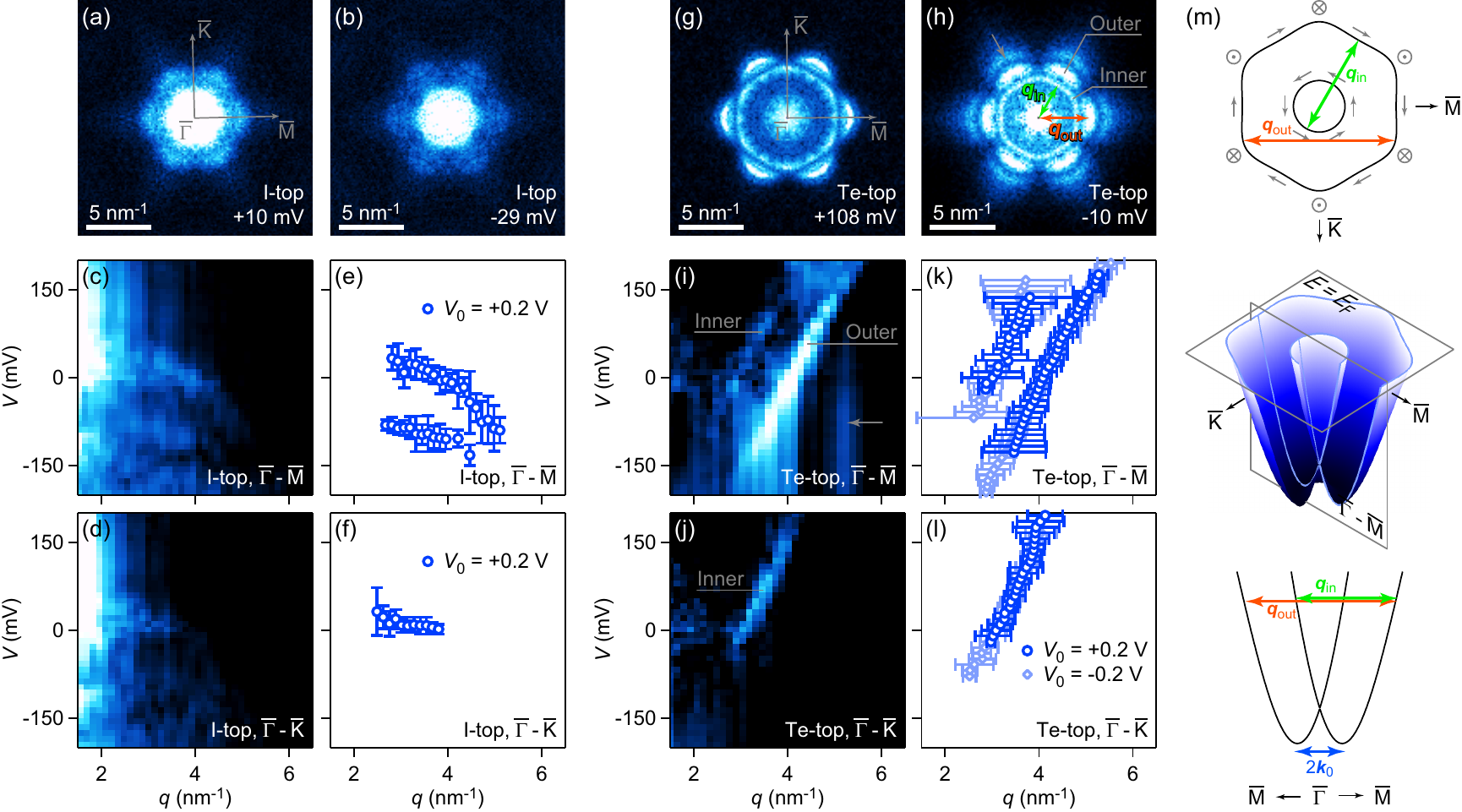}
	\caption{
		(Color online)
		Fourier analysis of the electron standing waves.
		(a, b, g, h) Fourier transforms of d$I$/d$V$ images.
		The setup bias voltages are -0.2~V for (a) and (b), and 0.2~V for (g) and (h).
		The one-sided gray arrows in (h) and (i) indicate the extrinsic feature described in the text.
		(c, d, i, j) High-symmetry linecuts of the Fourier transforms, taken along $\overline{\Gamma}-\overline\mathrm{M}$ (c and i), and $\overline{\Gamma}-\overline\mathrm{K}$ (d and j).
		(e, f, k, l) Peak positions extracted from (c), (d), (i), and (j), respectively, as quantitative eye guides to show the dispersive branches.
		The markers and the error bars are peak positions and widths extracted by Lorentzian fits, respectively.
		The error bars also indicate fitting directions.
		Peak positions extracted from another dataset taken with a setup voltage of -0.2~V are also shown in (k) and (l), demonstrating that the dispersive branches are independent of setup bias voltages.
		(m) Schematic figures of the surface band structure of the Te-top domain observed by ARPES: from top to bottom, the Fermi surface, a 3D illustration of the band structure, and the dispersion along $\overline{\Gamma}-\overline\mathrm{M}$.
		The gray arrows and markers denote directions in-plane and out-of-plane spin components.
		The latter component arises concomitantly with the hexagonal warping.
		$\bm{q}_\mathrm{in}$ and $\bm{q}_\mathrm{out}$ are scattering vectors of the observed electron standing waves.
		$\bm{k}_0$ is the momentum offset of the spin-split bands.
	\label{fig:QPI}}
\end{figure*}

To unambiguously conclude local carrier polarity of each domain, we focus on the dispersion relationships of the electron standing waves by analyzing Fourier transforms of d$I$/d$V$ images as a function of bias voltages.
In the I-top domain, a branch fans out from the $\overline{\Gamma}$ point with decreasing bias voltages as shown in Fig.~\ref{fig:QPI}(a) and \ref{fig:QPI}(b)~\cite{SMFFTmovie}.
The high-symmetry linecuts shown in Fig.~\ref{fig:QPI}(c)-\ref{fig:QPI}(f) exhibit a branch with negative slope (d$|\bm{q}|$/d$V < 0$) and crossing 0~mV as well as one more branch in deeper energies.
Therefore, as implied by the d$I$/d$V$ spectrum shown in Fig.~\ref{fig:BiTeI}(e), a 2D hole gas is formed at the surface of the I-top domain.
Meanwhile in the Te-top domain, the dispersion relationship is completely different from that in the I-top domain.
As shown in Fig.~\ref{fig:QPI}(g) and \ref{fig:QPI}(h), two dispersive branches are observed.
The outer branch appears as a peak in the $\overline{\Gamma}-\overline\mathrm{M}$ direction and the inner one as a hexagonal ring surrounding the $\overline{\Gamma}$ point.
Both branches approach to the $\overline{\Gamma}$ point with decreasing bias voltages~\cite{SMFFTmovie}.
The high-symmetry linecuts shown in Fig.~\ref{fig:QPI}(i)-\ref{fig:QPI}(l) exhibit that these branches have positive slope (d$|\bm{q}|$/d$V > 0$) and cross 0~mV.
Therefore, the surface carriers of the Te-top domain are electrons.
An energy-independent feature appearing on the outside of the outer branch is an extrinsic feature inherent to SI-STM because its location changes depending on setup bias voltages~\cite{Kohsaka2007}.
We note that the identification of the top layer and corresponding dispersion relationships are consistent with the ARPES results~\cite{Landolt2012}.

The clear electron standing waves in the Te-top domain bear closer analyses to unveil the nature of the electronic state.
The presence of two branches indicates that two scattering channels connecting non-orthogonal electronic states are involved with the electron standing waves~\cite{Petersen2000, Pascual2004, Roushan2009}.
By comparing the observed dispersions with the ARPES results~\cite{Ishizaka2011, Crepaldi2012, Landolt2012, Sakano2013}, we assign the scattering channels in the Te-top domain as illustrated in Fig.~\ref{fig:QPI}(m).
The inner branch arises from inter-band scattering between the spin-split bands as had ever been observed for Rashba-split states~\cite{Hirayama2011, El-Kareh2013}.
The outer branch originates from intra-band scattering in the hexagonally-warped outer band.
This branch emerges, as known for the surface Dirac Fermions of topological insulators~\cite{Zhang2009, Fu2009, Alpichshev2010, Beidenkopf2011}, because of a non-zero out-of-plane component of the spin polarization that is characteristic of $C_\mathrm{3v}$-symmetry of the crystal lattice~\cite{Ishizaka2011, Bahramy2012}.
Given this assignment of the two branches, we can estimate the momentum offset of the spin-split bands at ($\bm{q}_\mathrm{out}-\bm{q}_\mathrm{in})/2\sim 0.5$~nm$^{-1}$.
This value agrees well with the ARPES results~\cite{Ishizaka2011, Crepaldi2012, Landolt2012, Sakano2013}.
The spin splitting is observable in the electron standing waves of spin-split bands (without scattering to additional bands\cite{Steinbrecher2013,Leicht2014}) when the band structure deviates from a simple Rashba model of a theoretical prediction~\cite{Petersen2000} and is hexagonally warped.

\begin{figure*}
	\includegraphics{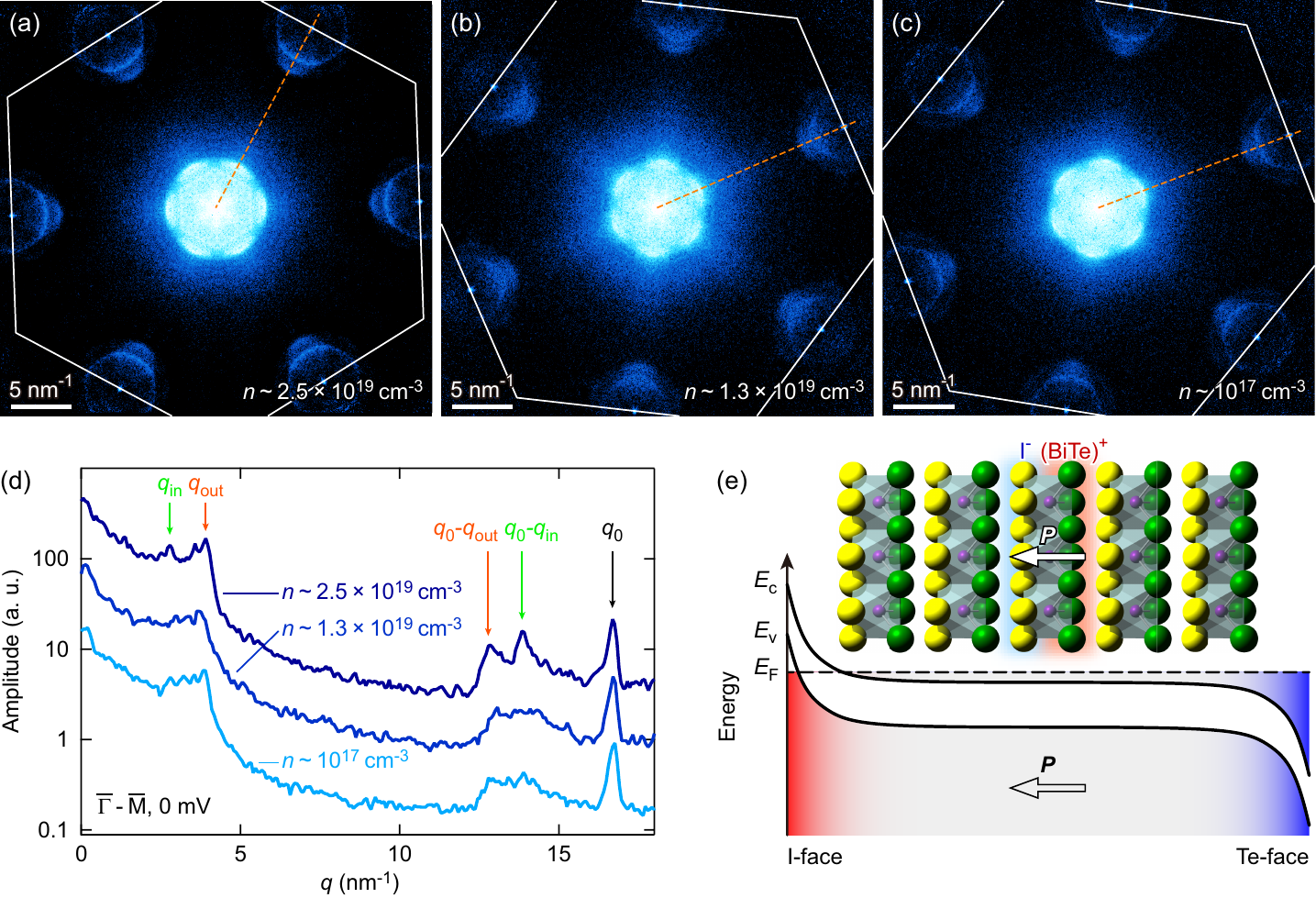}
	\caption{
		(Color online)
		$n$-independence of the 2D electron gas induced by the spontaneous electric polarization.
		($n$ is bulk electron density.)
		(a-c) Fourier transforms of d$I$/d$V$ images taken at 0~mV with a lock-in modulation of 1~mV$_\mathrm{rms}$ and a setup bias voltage of -20~mV.
		The solid white lines enclose the doubled $q$-space Brillouin zone where scattering inside the first Brillouin zone of $k$-space is found.
		The broken orange lines are trajectories along which the linecuts shown in (d) were taken.
		(d) $\overline{\Gamma}-\overline\mathrm{M}$ linecuts of the Fourier transforms shown in (a)-(c).
		Each curve is shifted vertically for clarity.
		$q_\mathrm{in}$, $q_\mathrm{out}$, and $q_0$ indicate positions of the inner branch, the outer branch, and the Bragg peak, respectively.
		The peaks annotated as $q_0$-$q_\mathrm{out}$ and $q_0$-$q_\mathrm{in}$ are replicas of the outer and inner branches, respectively.
		(e) A band diagram along the $c$-axis of BiTeI and a corresponding schematic of the crystal structure.
		$E_\mathrm{c}$, $E_\mathrm{v}$, and $E_\mathrm{F}$ are the bottom of conduction band, the top of valence band, and the Fermi level, respectively.
		The arrows annotated with ``$\bm{P}$'' indicate the direction of the spontaneous electric polarization.
	\label{fig:independent}}
\end{figure*}

The most salient feature of the electron standing waves in the Te-top domain is found in relationship to bulk electron density, $n$.
As shown in Fig.~\ref{fig:independent}(a)-\ref{fig:independent}(d), the Fourier transforms of d$I$/d$V$ images reveal that the wavevectors of two branches are independent of $n$.
This feature was commonly observed in the samples studied and thus is also independent of details of the domain structures.
The constant wavevector means that, as the Fermi level shifts as $n$ changes ($\sim$0.1~eV for the $n$ range studied~\cite{Lee2011}), the surface band also shifts such that the Fermi wavelength of the surface 2D state stays constant.
As indicated by the constant Fermi wavelength, the electron densities at the surface stay constant, even though those in the bulk change by two orders of magnitude.
The surface electron density is roughly estimated at $(\bm{q}_\mathrm{in}/2)^2/(2\pi) \sim 3\times 10^{13}$ cm$^{-2}$.
The peculiar $n$-independence is essential for specifying the mechanism of the 2D carriers as discussed below.

\begin{figure*}
	\includegraphics{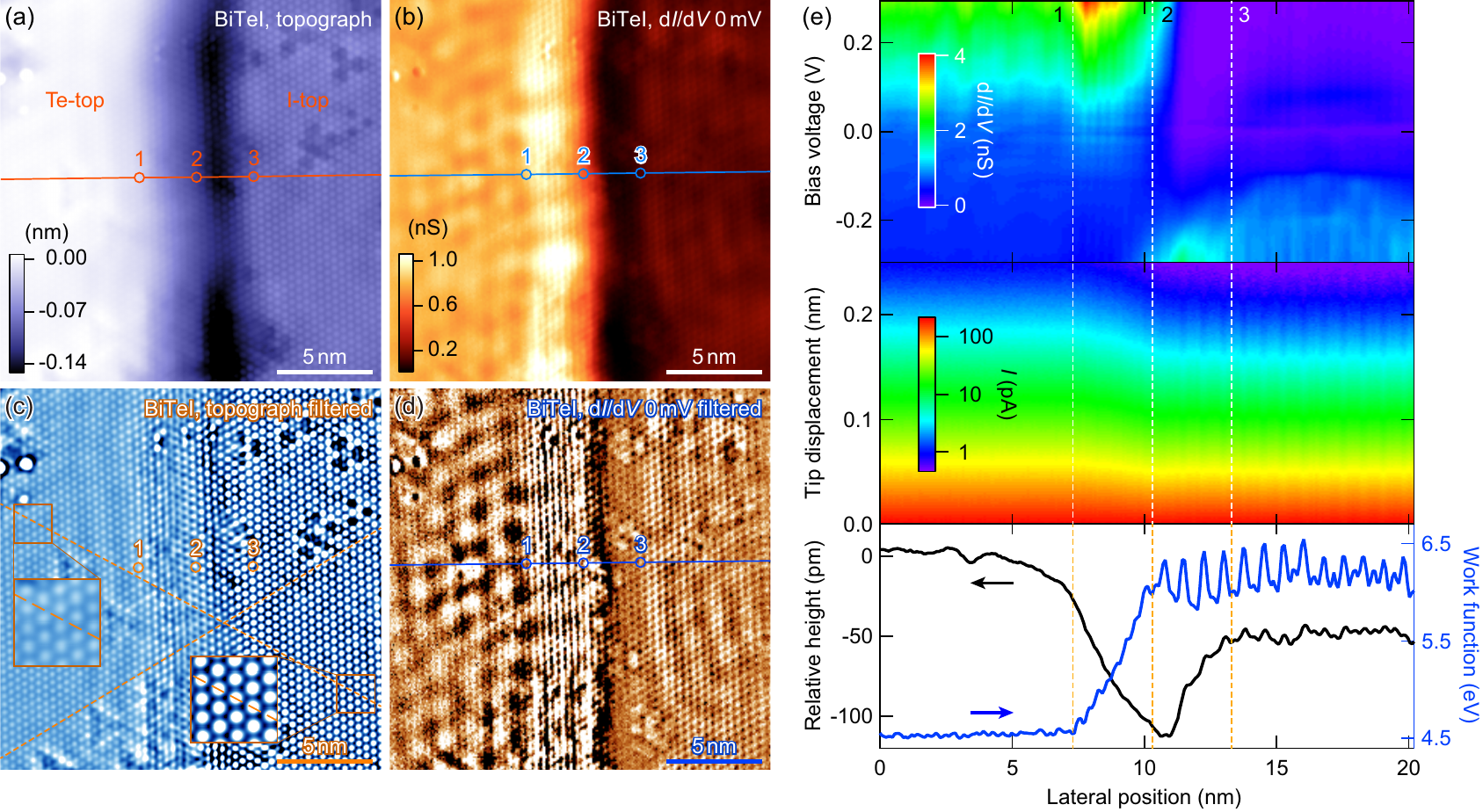}
	\caption{
		(Color online)
		Electronic structures of a lateral $p$-$n$ junction of the 2D carriers.
		(a) A 20 $\times$ 20~nm$^2$ topographic image around a domain boundary.
		The scanning parameters are 0.2~nA at -0.1~V.
		(b) A d$I$/d$V$ image taken in the same area as (a) at 0~mV with a setup bias voltage of -0.3~V.
		(c and d) Fourier-filtered images of (a) and (b), respectively.
		Low-wavenumber components are suppressed to enhance contrast of atomic corrugations.
		The broken lines are a guide to eyes for showing lateral shift between the topmost sublattices of the two domains.
		The insets are zoom-in images in the areas of the boxes.
		(e) Intensity plots of d$I$/d$V$ (top) and $I$-$z$ (middle) measurements taken along the trajectory shown as solid lines in (a), (b), and (d).
		A setup bias voltage of -0.3~V and a lock-in modulation of 4~mV$_\mathrm{rms}$ are used for the d$I$/d$V$ measurement.
		The bottom panel shows the corresponding profile of (a) and the work function calculated from the $I$-$z$ data.
		The positions 1 and 3 are edges of the depletion layer, and the position 2 is the middle point between 1 and 3.
		These positions are indicated as the circles in (a)-(d) and the broken lines in (e).
	\label{fig:pn}}
\end{figure*}

The electron standing waves are also observed around the domain boundary, providing details of the $p$-$n$ junction at the boundary.
Fig.~\ref{fig:pn}(a) and \ref{fig:pn}(b) show a topographic and a d$I$/d$V$ image taken in the same area around a domain boundary~\cite{SMdIdVmovie}.
To highlight atomic-scale structures, we apply a Fourier filter suppressing long-wavelength structures.
A structural model of the boundary is suggested from a filtered topographic image shown in Fig.~\ref{fig:pn}(c).
(See Appendix~\ref{appendix:model} for details.)
The electron standing waves in both domains are better visualized in a filtered d$I$/d$V$ image shown in Fig.~\ref{fig:pn}(d)~\cite{SMdIdVmovie}, indicating that this $p$-$n$ junction is a tunnel diode between highly-doped semiconductors.
The depletion layer of the $p$-$n$ junction manifests itself as the intervening zone without electron standing waves in Fig.~\ref{fig:pn}(d).

Transition of the electronic states across the $p$-$n$ junction is summarized in Fig.~\ref{fig:pn}(e).
d$I$/d$V$ spectral variation occurs in the intervening zone, defining the depletion layer.
The width of depletion layer is about 6~nm, in agreement with a simple estimate.
(See Appendix~\ref{appendix:width} for details.)
The depression observed in the topographic image corresponds to the depletion layer.
The work function has three characteristics; it has larger value in the I-top domain, shows larger modulations in the I-top domain, and changes solely in the Te-top domain side of the depletion layer.
The larger value is related to the origin of the ambipolar 2D carriers as described later as well as apparent heights of the domains (Appendix~\ref{appendix:apparentheight}). 
Further analysis is required for the larger modulations and the asymmetric change.

\section{Discussion}

How can the ambipolar 2D carriers be induced at the surface of BiTeI?
The key ingredient to specify the origin is the peculiar constant density of the 2D electrons.
Similar constant density is reported for SrTiO$_3$ where the 2D electrons are attributed to surface oxygen vacancies~\cite{Santander-Syro2011}.
In this case, the density of oxygen vacancies at the surface is much larger than and virtually independent of that in the bulk.
In contrast, however, extrinsic surface modifications are not observed in our samples (Appendix~\ref{sec:nodefect}) and therefore excluded as the origin of the 2D carriers.
Instead, the constant density in BiTeI is naturally explained by the spontaneous electric polarization in the bulk.
The unit BiTeI layer consists of a positively-charged (BiTe)$^+$ bilayer and a negatively-charged I$^-$ layer~\cite{Shevelkov1995}.
The spontaneous electric polarization therefore directs from a BiTe bilayer to an I layer within the unit layer.
The conduction (valence) band then is bent negatively (positively) and split off to form a 2D electron (hole) gas at the surface of the Te-top (I-top) domain, as illustrated in Fig.~\ref{fig:independent}(e).
Since the density of accumulated 2D electrons is determined by strength of the polarization, it is independent of bulk electron density.

The other observations are also comprehensively understood by the spontaneous electric polarization.
The work function larger in the I-top domain is consistent with stronger confinement of electrons into the bulk caused by the spontaneous electric polarization pointing from a BiTe bilayer to an I layer.
The strong scattering at Bi-site (Te-site) in the Te-top (I-top) domain reflects the orbital character of the surface band split from the conduction (valence) band where Bi-6p (Te-5p) orbitals predominantly contribute~\cite{Bahramy2012}.
The domain structure with the $\overline{2}$ relationship and the meandering boundaries is analogous with those of displacive-type ferroelectrics.
We thus conclude that the ambipolar 2D carriers at BiTeI surfaces are induced by the spontaneous electric polarization in the bulk.

In this study, we have precisely determined local carrier polarity by observing electron standing waves with SI-STM.
This results demonstrates that, to probe carrier polarity, SI-STM is available complementarily to other techniques such as ARPES, thermoelectric probes, single electron transistors, and photoelectron emission microscopy.
Moreover, we unveiled that the underlying mechanism of electron standing waves is common to that of Dirac Fermions in topological insulators.
This enables us to examine Rashba-type spin splitting with SI-STM and provides a unified framework to address spin-dependent scattering phenomena that are a key aspect for spintronics applications.
Most importantly, our study establishes that the spontaneous electric polarization induces ambipolar 2D carriers.
Carrier densities of these 2D carriers may be controlled by strain and temperature via the piezoelectric and pyroelectric effects.
Since the $p$-$n$ junction at the domain boundary is a consequence of the ambipolar 2D carriers and the domain structure sharing the common root, the surface of a polar material is a new platform to investigate a lateral $p$-$n$ junction of 2D carriers.
Such a surface can be a potential substructure to study unconventional devices such as a topological $p$-$n$ junction~\cite{Wang2012}.
The spontaneous electric polarization is, as a means to induce 2D carriers, complementary to field effect transistors and heterostructures in the sense that the induced carriers are non-volatile and polarity-switchable, and thus is available to explore new phenomena and functionalities.

\begin{acknowledgments}
We thank K. Ishizaka and M. Kawamura for fruitful discussions.
This work was supported by JSPS KAKENHI Grant numbers 24340078 and 24684022.
\end{acknowledgments}

\appendix

\section{Bias dependence of relative heights}
\label{appendix:relativeheight}
\begin{figure}
	\includegraphics[keepaspectratio]{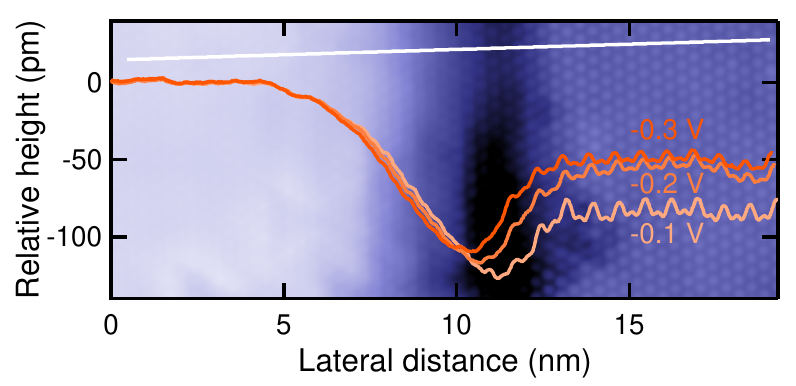}
	\caption{
		(Color online)
		Bias dependence of apparent height.
		The orange curves are line profiles taken across a boundary of the two types of domains.
		The background image shows a 20~nm $\times$ 8~nm topographic image taken at -0.1~V around the boundary.
		The white line denotes a trajectory where the profiles were taken.
	}
	\label{fig:DBtopo}
\end{figure}

Apparent heights of the two types of domains depend on the bias voltages as shown in Fig.~\ref{fig:DBtopo}, indicating that the electronic states at the surfaces of the domains are different and the electronic difference contributes to the difference of apparent heights.
This is contrasting to a step where its height is independent of the bias voltages.
See also Appendix~\ref{appendix:apparentheight} about contributions to the apparent height.

\section{Domains with opposite stacking sequences}
\label{appendix:stacking}
\begin{figure}
	\includegraphics[keepaspectratio]{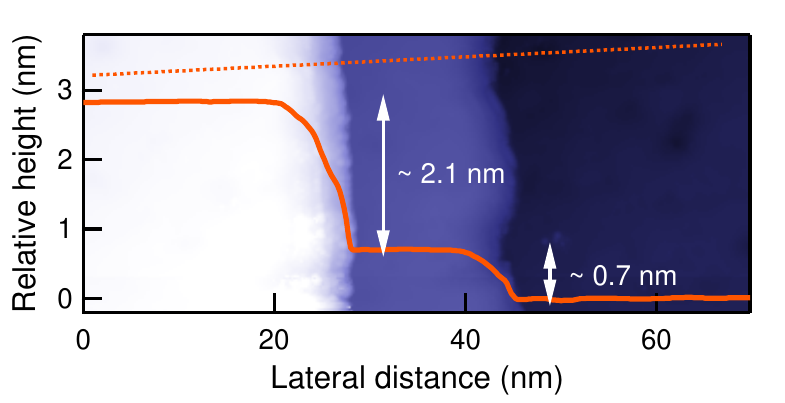}
	\caption{
		(Color online)
		Atomic steps with heights of multiples of the $c$-axis constant.
		The solid curve is a line profile taken across atomic steps.
		The background image shows a 74~nm $\times$ 31~nm topographic image taken at -0.2~V.
		The dotted line denotes a trajectory where the profile was taken.
	}
	\label{fig:steptopo}
\end{figure}

The experimental fact that {\it only two} kinds of domains are observed means that number of surface structures is two.
If there are multiple cleavage planes in the ideal crystal structure, more than two kinds of surface structures appear due to combinations of the topmost and subsurface layers.
Stacking faults also give more than two kinds of surface structures for the same reason.
Multiple cleavage planes and stacking faults are excluded also by the step heights that are multiples of the $c$-axis lattice constant (Fig.~\ref{fig:steptopo}).
Then, the topmost layer of each domain must be one of the two layers adjacent to the natural cleavage plane, and the second-topmost layer must be one remaining layer.
Namely, Te and I layers are the top and a Bi layer is the second.
That is, the observed domain structures are composed of opposite stacking sequences, Te-Bi-I (Te-top) and I-Bi-Te (I-top).
This domain structure naturally accounts for that spin polarization is observed to be unchanged even when a crystal is flipped~\cite{Landolt2012}, as suggested by the authors.

\section{Details about the domain structure}
\label{appendix:domaindetails}

\subsection{Identification of stacking sequence}
\begin{figure*}
	\includegraphics[keepaspectratio]{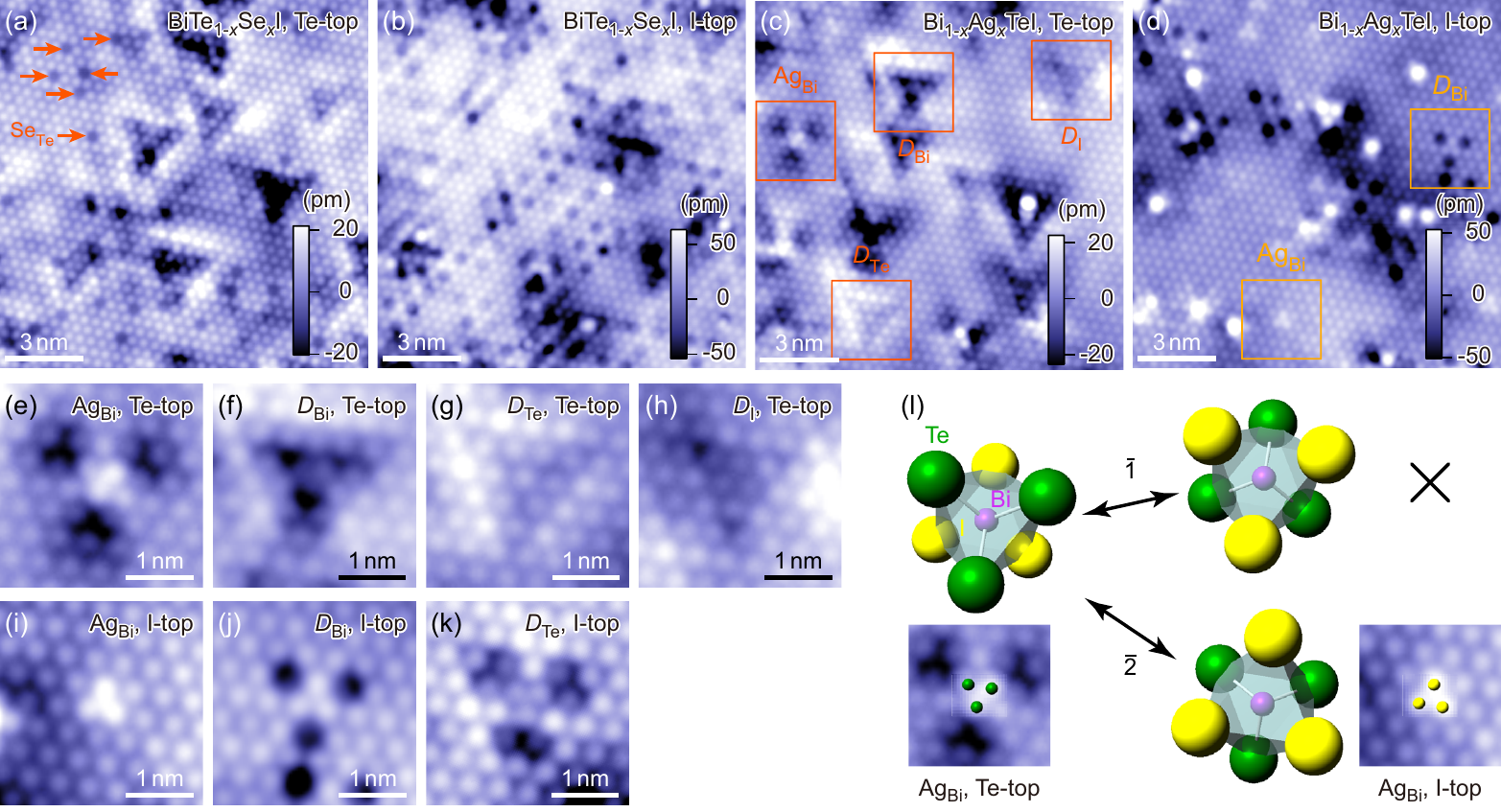}
	\caption{
		(Color online)
		Topographic images of substituted samples to identify the domain structure.
		(a)-(d) 14$\times$14~nm$^2$ topographic images of BiTe$_{1-x}$Se$_x$I ((a) and (b)) and Bi$_{1-x}$Ag$_x$TeI ((c) and (d)).
		Setup bias voltages are -20~mV (a), -0.4~V ((b) and (d)), and -0.2~V (c).
		For each sample, the images were taken on a single surface.
		Arrows in (a) indicate several exemplary Se atoms substituted for Te atoms.
		The annotations $D_\mathrm{X}$ denote defects at X-site (X = Bi, Te, and I).
		(e)-(k) Close-up images around impurities and defects in 3$\times$3~nm$^2$ squares.
		Areas of (e)-(h) are shown as boxes in (c) and (d).
		(k) is an image around $D_\mathrm{Te}$ indicated by the arrow in Fig.~\ref{fig:BiTeI}(c).
		This image is rotated to fit the other images.
		(l) Point group symmetries to realize opposite stacking sequences.
		Schematic figures of the topmost Te and I atoms are superimposed on topographic images of Ag$_\mathrm{Bi}$ clipped from (e) and (i).
		Location of Ag$_\mathrm{Bi}$ relative to the topmost sublattice proves that the structural relationship between the domains is $\overline{2}$ (reflection).
}
	\label{fig:impurities}
\end{figure*}

To identify stacking sequences of the domains, we measured BiTe$_{1-x}$Se$_x$I where Se substituted for Te works as a marker of Te site.
As shown in Fig.~\ref{fig:impurities}(a) and \ref{fig:impurities}(b), Se atoms are observed at the atomic site of Fig.~\ref{fig:impurities}(a), indicating that this type of domain with dark triangles has the Te-top stacking and the other with three dots has the I-top stacking.

\subsection{Point-group operation between the domains}
Here we consider the structural relationship between the two types of domains in terms of a point-group operation.
As is evident from two topographic images taken on a single surface (Fig.~\ref{fig:impurities}(a) and \ref{fig:impurities}(b), and Fig.~\ref{fig:impurities}(c) and \ref{fig:impurities}(d)), the orientation of the topmost sublattices is identical for the two types of domains.
Point-group operations to realize opposite stacking sequences with keeping orientation of the topmost sublattices are $\overline{1}$ (inversion), $\overline{2}$ (reflection), $\overline{3}$, and $\overline{6}$.
Actually, only $\overline{1}$ and $\overline{2}$ are independent because of the three-fold crystal symmetry of BiTeI.

A distinction between $\overline{1}$ and $\overline{2}$ is given by locations of Bi atoms.
With respect to the topmost sublattice, there are two possible sites where Bi atoms can occupy.
If $\overline{1}$ is the case, Bi atoms occupy the same site in both domains.
If $\overline{2}$ is the case, they occupy different sites.
To make this distinction, we measured topographic images of Bi$_{1-x}$Ag$_x$TeI as shown in Fig.~\ref{fig:impurities}(c) and \ref{fig:impurities}(c).
Since Ag$_\mathrm{Bi}$ atoms are observed at different sites (Fig.~\ref{fig:impurities}(e)and \ref{fig:impurities}(i)), the point-group operation is $\overline{2}$ (reflection).
The above discussion is summarized in Fig.~\ref{fig:impurities}(l).
This successful identification of the domain structure demonstrates that a combination of crystal growth and scanning tunneling microscopy works complementarily to diffraction techniques and transmission electron microscopy to determine local structures.

\subsection{Identification of defect patterns}
Fig.~\ref{fig:impurities}(e)-\ref{fig:impurities}(k) show topographic patterns of defects and impurities.
Identifying a defect pattern centered at the topmost Te atoms is straightforward (Fig.~\ref{fig:impurities}(g)).
Although defects bright in color are often found at the topmost I site, we have not yet identified a feature common to all images of the I-top domain.
Bi-site defects (Fig.~\ref{fig:impurities}(f) and \ref{fig:impurities}(j)), found at an interatomic site, are identified based on locations of Ag$_\mathrm{Bi}$ (Fig.~\ref{fig:impurities}(e) and \ref{fig:impurities}(i)).
Patterns centered at the other interatomic site are identified as remaining ones, I-site defects for the Te-top domain (Fig.~\ref{fig:impurities}), and Te-site defects for the I-top domain (Fig.~\ref{fig:impurities}(k)).

\section{A structural model of the domain boundary}
\label{appendix:model}
\begin{figure}
	\begin{center}
		\includegraphics{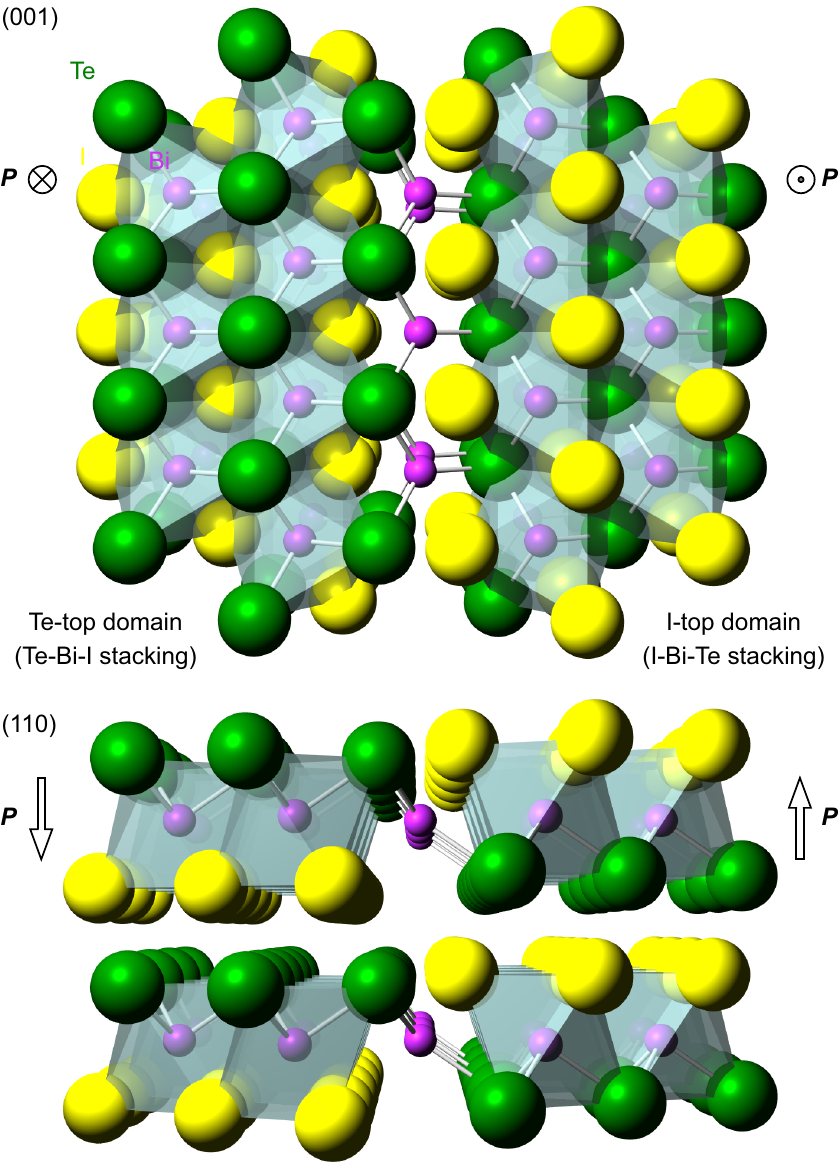}
	\end{center}
	\caption{
		(Color online)
		A structural model of the domain boundary, seen from the top and the side.
		The arrows annotated with ``$\bm{P}$'' indicate directions of the spontaneous electric polarization.
	}
	\label{fig:model}
\end{figure}

Supplementary Figure~\ref{fig:model} shows a structural model of the domain boundary.
This model is constructed to satisfy the $\overline{2}$ relationship between the domains and the lateral shift of the topmost sublattices shown in Fig.~\ref{fig:pn}(c).
As for a possible vertical shift of the domains, we can not separate a morphological height difference from the apparent height difference shown in the topographic image (Fig.~\ref{fig:pn}(a)) because the LDOS and the work function also contribute to the apparent height.
(See also Appendix~\ref{appendix:apparentheight} for a mathematical description.)
The Bi sublattice, therefore, is drawn as flat for simplicity.

\section{Width of the depletion layer}
\label{appendix:width}
Here we calculate the width of the depletion layer of a planer $p$-$n$ junction, $d$, with reasonable parameters to compare with the width of the observed depletion layer.
$d$ is given by
\begin{equation}
	d = \sqrt{\frac{2\epsilon_0\epsilon V_\mathrm{D}}{e}\frac{N_\mathrm{A}+N_\mathrm{D}}{N_\mathrm{A}N_\mathrm{D}}},
\end{equation}
where $\epsilon_0$ and $e$ are the vacuum permittivity and the electron charge, respectively.
Others are material-dependent parameters.
$\epsilon$ is the dielectric constant.
$V_\mathrm{D}$ is the built-in potential.
$N_\mathrm{A}$ and $N_\mathrm{D}$ are the density of acceptors and donors, respectively.
We use $N_\mathrm{A} = N_\mathrm{D} = {n_\mathrm{2D}}^{3/2}$ where $n_\mathrm{2D}$ is the 2D electron density on the Te-top domain.
$n_\mathrm{2D}$ is estimated by the Fermi wavenumber $k_\mathrm{F}$ and the wavenumber of the electron standing wave at 0~mV, $q_\mathrm{in}$.
Namely, $n_\mathrm{2D}={k_\mathrm{F}}^2/(2\pi)$ and $k_\mathrm{F}=q_\mathrm{in}/2$.
Since $q_\mathrm{in}\sim$ 2.8~nm$^{-1}$, we get $N_\mathrm{A} = N_\mathrm{D}\sim 1.7\times 10^{20}$ cm$^{-3}$.
Other parameters are $\epsilon\sim 19$~(Ref.~\onlinecite{Lostak1980}) and $V_\mathrm{D}\sim$ 1.6~V that is estimated from the difference of the work function shown in Fig.~\ref{fig:pn}(e).
By using these values, we get $d\sim 6.2$~nm.

\section{Apparent height in a topographic image}
\label{appendix:apparentheight}
Here we give a simple mathematical description about multiple contributions to apparent height of a constant-current topographic image.
The STM tunneling current $I$ is given by
\begin{equation}
	I(\bm{r},z,V) = C\exp(-2\kappa(\bm{r})z)\int_0^{eV}\!N(\bm{r},E)\mathrm{d}E
\end{equation}
while $z$ is the tip's surface-normal coordinate, $V$ is the sample bias, and $N(\bm{r},E)$ is the sample's LDOS at lateral locations $\bm{r}$ and energy $E$.
$\kappa$ is related to the tunneling work function $\kappa(\bm{r}) = \sqrt{2m\phi(\bm{r})}/\hbar$.
$e$, $m$, and $\hbar$ are the elementary charge, the electron mass, and the Planck constant, respectively.
Given this formula, a constant-current ($I_0$) topographic image $z_\mathrm{cc}(\bm{r};I_0,V_0)$ at a sample bias voltage $V_0$ is given by
\begin{equation}
	z_\mathrm{cc}(\bm{r};I_0,V_0) = \frac{\hbar}{\sqrt{8m\phi(\bm{r})}}\ln\left(\frac{C}{I_0}\int_0^{eV_0}\!N(\bm{r},E)\mathrm{d}E\right).
	\label{eq:z0}
\end{equation}
Eq.~(\ref{eq:z0}) indicates that apparent height of a constant-current topographic image is low when the work function is large or the LDOS is small.
This is a consequence of a feedback loop that gets a tip closer to a sample to compensate fast decay of the wave function or reduction of LDOS available for tunneling.
In the case of BiTeI, the work function is larger in the I-top domain as shown in Fig.~\ref{fig:pn}(e).
The LDOS is likely to be smaller in the I-top domain because the 2D hole state in the I-top domain is caused by strong inversion whereas the 2D electron state in the Te-top domain by accumulation.
Namely, both the work function and the LDOS contribute to the apparent height lower in the I-top domain.

\section{No extrinsic surface modification}
\label{sec:nodefect}
Here we describe that our samples have neither surface absorbates nor additional defects possibly changing carrier density at the surface.
As is evident in the topographic images shown in Fig.~\ref{fig:BiTeI}, no surface absorbate exists.
Surface defects additionally generated by cleavage are excluded by counting the density of defects as follows.

To evaluate the density of defects near the surface, we assume that observed defects lie in the top unit layer.
This assumption gives the highest estimate of defect density.
By counting defects in several topographic images, we estimate that the density of these defects is about $5-8\times 10^{19}$ cm$^{-3}$.
This value, slightly larger than the bulk electron density ($2.5\times 10^{19}$ cm$^{-3}$), is just as expected because a considerable amount of acceptors exists in this material as well as predominant donors~\cite{Horak1985}.
Therefore, the density of defects near the surface is virtually identical to that in the bulk without increased by cleavage.

\end{document}